\documentclass[review]{elsarticle}
\usepackage[utf8]{inputenc}
\usepackage[english]{babel}
\usepackage{amssymb}
\usepackage[review]{changes}
\usepackage{natbib}
\usepackage{filecontents}
\usepackage{lineno,hyperref}
\usepackage{amsthm, amsmath, amsfonts, txfonts, bm, dsfont}
\usepackage{booktabs, footnote}
\usepackage[utf8]{inputenc}
\usepackage{colortbl, xcolor}
\usepackage{verbatim}
\usepackage[english]{babel}

\usepackage{enumitem}
\newlist{steps}{enumerate}{1}
\setlist[steps, 1]{wide=0pt, leftmargin=\parindent, label=Step \arabic*:, font=\bfseries}

\modulolinenumbers[5]
\usepackage{floatrow}
\let\today\relax
\makeatletter
\def\ps@pprintTitle{%
	\let\@oddhead\@empty
	\let\@evenhead\@empty
	\def\@oddfoot{\footnotesize\itshape
		{Submitted to AMASES Award} \hfill\today}%
	\let\@evenfoot\@oddfoot
}
\makeatother
\newtheorem{theorem}{Theorem}

\theoremstyle{definition}
\newtheorem{definition}{Definition}[section]

\newtheorem{assumption}{Assumption}[section]

\begin{document}

\begin{frontmatter}

\title{A cohort-based Partial Internal Model for demographic risk}


\author[label1]{Gian Paolo Clemente}
\ead{gianpaolo.clemente@unicatt.it}
\address[label1]{Department of Mathematics for Economic, Financial and Actuarial Sciences\\ Università Cattolica del Sacro Cuore, Milano}

\author[label1]{Francesco Della Corte}
\ead{francesco.dellacorte1@unicatt.it}

\author[label1]{Nino Savelli}
\ead{nino.savelli@unicatt.it}




\begin{abstract}
We investigate the quantification of demographic risk in a framework consistent with the market-consistent valuation imposed by Solvency II. We provide compact formulas for evaluating inflows and outflows of a portfolio of insurance policies based on a cohort approach. In this context, we maintain the highest level of generality in order to consider both traditional policies and equity-linked policies: therefore, we propose a market-consistent valuation of the liabilities. In the second step we evaluate the Solvency Capital Requirement of the idiosyncratic risk, linked to accidental mortality, and the systematic risk one, also known as trend risk, proposing a formal closed formula for the former and an algorithm for the latter. We show that accidental volatility depends on the intrinsic characteristics of the policies of the cohort (Sums-at-Risk), on the age of the policyholders and on the variability of the sums insured; trend risk depends both on accidental volatility and on the longevity forecasting model used.
\end{abstract}

\begin{keyword}
\texttt{Demographic risk, Solvency II, Market-consistent valuation, Risk Theory, Cohort approach} 
\end{keyword}

\end{frontmatter}


\section{Introduction}\label{sec:intro}
In the recent years, there has been a trend towards market consistent valuation of assets and liabilities in the insurance field. Both Solvency II Directive \cite{solvency2} and IFRS accountings standard \cite{IFRS} defined a market-consistent valuation of technical liabilities. Market-consistent valuation indeed plays a crucial role in regulatory compliance and financial solvency assessment for insurance companies. Regulators require insurers to value their liabilities using market-consistent methods to ensure accurate and reliable financial reporting. \\
In particular, assessing demographic risk is an important aspect of market-consistent valuation for life insurance company. Demographic risk refers to the uncertainty and variability associated with demographic factors such as mortality rates, longevity, and morbidity. It is a crucial component in valuing insurance liabilities accurately, as these factors have a significant impact on the timing and amount of future policy benefits and claims obligations. The assessment of demographic risk involves incorporating appropriate demographic assumptions and models into the valuation process. These assumptions are typically based on historical data, expert judgment, and relevant population statistics. The aim is to capture the inherent variability and uncertainty in demographic factors to produce more realistic and reliable valuation outcomes. \\
In the literature, a common approach to assess capital requirement for demographic risk is through stochastic modelling and several contributions have been provided (see, e.g., \cite{boonen2017solvency, hari2008longevity, olivieri2008assessing, plat2011one, stevens2010calculating}). \\
Several papers focused also on market-valuation in life insurance. In particular, Dhaene et al. (see
\cite{dhaene2017fair}) introduce a fair valuation of liabilities related to a portfolio in a single period framework, such that it is both mark-to-market for any hedgeable part of a claim, and mark-to-model for any claim that is independent of financial market evolutions.
\cite{delong2019fair1} and \cite{delong2019fair2} focus instead on the market-consistent evaluation in continuous time. A valuation of some life insurance contracts in a stochastic interest rate environment taking into account the default risk of the underlying insurance company is given in \cite{BERNARD2005499}. \cite{GROSEN} consider typical participating policy showing that they can be decomposed into a risk free bond element, a bonus option, and a surrender option. A dynamic model is then constructed in which these elements can be valued separately using contingent claims analysis.  \cite{Moehr} propose a valuation framework based on replication over multiple 1-year time periods by a periodically updated portfolio of assets. \\
The mathematical framework that leads to market-consistent values for insurance liabilities is presented in detail in \cite{wuthrich2010market} and aspects related to Solvency topics are treated in \cite{wuthrich2013financial}. Approaches based on two and three steps procedure have been also provided for market-consistent valuation of liabilities. \cite{BArigou} combine quadratic hedging with application of a risk measure on the residual liability, to obtain a cost-of-capital margin. A three-step hedge-based valuation for the valuation of hybrid claims is given in \cite{Linders}.\\
However few attention has been paid to combine market-consistent valuation and the assessment of capital requirement for demographic risk. By assessing demographic risk within the framework of market-consistent valuation, insurance companies can better understand and manage their exposure to longevity risk, mortality risk, and other demographic uncertainties. This enables insurers to make informed decisions regarding pricing, reserving, and risk management strategies. \\
Therefore, we propose here a novel methodology for assessing capital requirement for mortality and longevity risk of linked policies. To this end, starting from the framework defined in \cite{wuthrich2010market} and \cite{wuthrich2013financial}, we consider a portfolio of homogenous policies and we provide a method for the evaluation of idiosyncratic and trend risks considering also financial effects. We show that accidental volatility is mainly related to the sum-at-risk of the considered contracts, the age of the policyholders and the variability of the insured sums. Trend risk is instead related to the accidental volatility and to the volatility of the model used for forecasting future longevity behaviour. Furthermore, we use a matrix approach that allows us to simulate a large number of scenarios in an extremely short time, in order to obtain a punctual and robust estimate of the risk-based capital. \\
A numerical application is developed in order to provide additional insights about main drivers that affect capital requirement. \\
The paper is organized as follows. In Section 2 we introduce some preliminary aspects of the model. In Section 3, we outline the mathematical framework of the model and in Section 4, we present the Cohort Valuation Portfolio using a matrix approach. Moving on to Section 5, we introduce the stochastic model designed to identify and quantify idiosyncratic and trend risks. Lastly, in Section 6, we provide a numerical example with a specific emphasis on the assessment of the Solvency Capital Requirement in line with Solvency II rules.
\newpage

\section{Preliminaries}
We indicate with $\Omega$ a given set, then with $\mathcal{F}$ a family of subsets of $\Omega$  with usual desired properties (see e.g., \cite{oksendal2003stochastic}), hence $\mathcal{F}$ is a $\sigma$-algebra and $(\Omega, \mathcal{F})$ is a measurable space.\\
Considering a generic fixed positive number $n$, we assume that for each $t\in[0,n]$ there is a $\sigma$-algebra; the collection of $\sigma$-algebras $\mathcal{F}_t$ is a filtration, denoted with $\mathbb{F}$. We define with $\emph{P}$ the probability measure that, to every set $\emph{A}\in \mathcal{F}$ assigns a number in $[0,1]$ (i.e., $\emph{P}(\emph{A})$) coherently with real-word evidence.\\
The starting point coincides with the definition of the vector of the cashflows, generally indicated with $\textbf{X}$ \footnote{Random variables are indicated with capital letters, while deterministic values with small letters}
\begin{equation}
	\textbf{X}=(\emph{X}_0,\emph{X}_1,...,\emph{X}_n)\in \emph{L}^2_{n+1}(\emph{P},\mathbb{F})
	\label{eq:cashflows}
\end{equation}
where $\emph{L}^2_{n+1}(\emph{P},\mathbb{F})$ is a Hilbert space with scalar product $\langle\cdot,\cdot\rangle$:
\begin{align}
	\emph{E}\left[\sum_{t=0}^{n}X^2_t\right]<\infty &\text{\space\space\space for all $\textbf{X}\in\emph{L}^2_{n+1}(\emph{P},\mathbb{F})$}
	\label{eq:hilbert_1}
\end{align}
\begin{align}
	\emph{E}\left[\langle\textbf{X},\textbf{Y}\rangle\right]=\emph{E}\left[\sum_{t=0}^{n}X_t\cdot Y_t\right]<\infty&\text{\space\space\space for all $\textbf{X},\textbf{Y}\in\emph{L}^2_{n+1}(\emph{P},\mathbb{F})$}
	\label{eq:hilbert_2}
\end{align}
\begin{align}
	\vert\vert\textbf{X}\vert\vert= {\langle\textbf{X},\textbf{X}\rangle}^{1/2}<\infty&\text{\space\space\space for all $\textbf{X}\in\emph{L}^2_{n+1}(\emph{P},\mathbb{F})$}
	\label{eq:hilbert_3}
\end{align}
For several purposes, contracts are grouped in cohorts in the actuarial practice (see, e.g., market consistent embedded value, IFRS, solvency capital requirement). Therefore, we consider here the following definition of cohort.
\begin{definition}[\textit{Cohort}]
\label{def:coorte}
We define a cohort a set of policyholders that have underwritten the same contract and that have the same characteristics. The only difference inside a single cohort regards the sums insured.  It is therefore assumed that the survival (and death) of individual policyholders is independent from each other and that all those who belong to the same cohort have the same survival probability.
\end{definition}
\noindent At the inception date (i.e., $ t = 0 $), we consider a number of $l_{0}$ policyholders in the cohort and we denote with $\textbf{s}_0=[s_{k,0}]$ ($k=1,2,...,l_{0}$) the vector of the sums insured of the policyholders. At the inception, each policyholder buys $s_{k,0}$ shares of the portfolio by paying the necessary premium to build the replicating portfolio. 
At each time period $t$ (with $t>0$), the information available is represented by $\mathcal{F}_t$. Therefore, the stochastic vector of the sums insured at time $t$ is defined as follows:
\begin{equation}
	\textbf{S}_t = \begin{pmatrix}
		s_{1,0} \\   s_{2,0} \\ ...  \\   s_{l_0,0}
	\end{pmatrix} \circ \begin{pmatrix}
		\mathbb{I}_{1,0}^L \\   \mathbb{I}_{2,0}^L \\ ...  \\   \mathbb{I}_{l_0,0}^L
	\end{pmatrix}\circ...\circ\begin{pmatrix}
		\mathbb{I}_{1,t-1}^L \\   \mathbb{I}_{2,t-1}^L \\ ...  \\  \mathbb{I}_{l_0,t-1}^L
	\end{pmatrix}
	\label{insured_sum_t}
\end{equation}
where with \lq\lq$\circ$" we indicate the Hadamard product  (also known as element-wise product) and the generic $\mathbb{I}_{k,t-1}^L$ is a dichotomic $\mathcal{F}_t$-measurable random variable. It assumes the value 1 if the $k$-th policyholder survived in $(t-1,t]$, 0 otherwise.

\begin{assumption}[\textit{Valuation functional and deflators}]
	\label{valfun}
	We assume that $\mathcal{Q}$ : $\emph{L}^2_{n+1}(\emph{P},\mathbb{F}) \rightarrow \mathbb{R}$ is a linear, positive, continuous and normalized functional on $\emph{L}^2_{n+1}(\emph{P},\mathbb{F})$ (hence, $\mathcal{Q}$ is a valuation functional).\\
	For more information on the valuation functional $\mathcal{Q}$, we refer to \cite{buhlmann1992stochastic} and \cite{buhlmann2000life}. However, we point out that $\mathcal{Q}_t(\textbf{X})$ is a monetary value, $\mathcal{F}_t$-measurable and therefore stochastic at time $0$.\\
	Taking advantage of the Riesz–Fréchet representation theorem (see Theorem 2.5 in \cite{wuthrich2010market}), there exists a $\bm{\varphi}\in\emph{L}^2_{n+1}(\emph{P},\mathbb{F})$ such that for all $\textbf{X}\in\emph{L}^2_{n+1}(\emph{P},\mathbb{F})$ we have
	\begin{equation}
		\mathcal{Q}_{0}\left(\textbf{X}_{(0)}\right)=\dfrac{1}{\varphi_0}\emph{E}\left[\sum_{t=0}^{n} \varphi_t X_t \right]=\emph{E}\left[\langle \left(\bm{\varphi}_{(0)}\right)^\intercal,\bm{X}_{(0)}\rangle\right]
		\label{eq:deflator2}
	\end{equation}
	Notice that we use the notation $\textbf{X}_{(t)}$ to indicate a generic vector $\textbf{X}$ which elements are $\left(X_t\right)_{t,t+1,...,n}$\footnote{This notation is taken from \cite{wuthrich2010market}}.
	
	\noindent The vector $\bm{\varphi}$ is the state price deflator, with the usual properties to be $\mathbb{F}$-adapted and $(\emph{P}, \mathbb{F})$-martingale, it has square integrable components, it is normalized ($\varphi_0=1$), its components are strictly greater than 0 and, whereas the market is complete, it's unique.\\
	Since the deflators assign market-consistent values to the cashflows $X_h$, with $h\in[1,n]$, it is possible to represent the reserves at time $t$, with $t\in[0,h-1]$ for outstanding cashflows as
	\begin{equation}
		\label{generic_reserve}
		\mathcal{R}_t^{(h)}=\mathcal{Q}_t\left(\textbf{X}_{(h)}\right)=\dfrac{1}{\varphi_t}\emph{E}\left[\sum_{s=h}^{n}\varphi_s X_s\bigg| \mathcal{F}_t\right]
	\end{equation}
\end{assumption}

\begin{definition}[\textit{Fair value principle}]
\label{Def:FV}
\noindent Article 76 of Directive 2009/138/EC (usually called Solvency II, see \cite{solvency2}) requires that the insurance and reinsurance undertaking hold technical provisions measured at fair value, i.e., \lq\lq\textit{the current amount insurance and reinsurance undertakings would have to pay if they were to transfer their insurance and reinsurance obligations immediately to another insurance or reinsurance undertaking}".\\
In literature, many authors have dealt with the topic in life insurance. \textit{Ante litteram}, Brennan \& Schwartz (see \cite{brennan1976pricing}) evaluated the equilibrium pricing of equity-linked life insurance policies with an asset value guarantee. Dhaene et. al. (see \cite{dhaene2017fair}), investigated the fair value of liabilities intended as both a market-consistent valuation for hedgeable claims (or parts of them), and actuarial for non-hedgeable claims. Delong et. al. (see \cite{delong2019fair1} and \cite{delong2019fair2}), extended the market consistent and actuarial valuation to the continuous time. We assume here that $\mathcal{Q}$ is a valuation functional (for further details see Definition \ref{valfun} or \cite{buhlmann1992stochastic}) hence, for a given state price deflator $\bm{\varphi}$, the process $\left\{\mathcal{Q}_t(\textbf{X})\right\}_{t=0,...,n}$ is consistent with an arbitrage-free valuation scheme.\\
In conclusion, we highlight that $\bm{\varphi}$-consistency implies that $\left(\varphi_t\mathcal{Q}_t\left(\textbf{X}_h\right)\right)_{t=0,...,n}$ is a $\left(\emph{P},\mathbb{F}\right)$-martingale.
\end{definition}

\begin{assumption}[\textit{Independent split of filtrations}]
	As in \cite{wuthrich2010market} and \cite{wuthrich2013financial}, we factorize the filtered probability space $(\Omega, \mathcal{F}, \emph{P}, \mathbb{F})$ into a product space to obtain an independent decoupling:
	\begin{equation}
		\mathbb{T}=\left(\mathcal{T}_t\right)_{t=0,...,n}
		\label{eq:T}
	\end{equation}
	\begin{equation}
		\mathbb{G}=\left(\mathcal{G}_t\right)_{t=0,...,n}
		\label{eq:G}
	\end{equation}
	where formula (\ref{eq:T}) refers to the filtration of insurance technical events and formula (\ref{eq:G}) refers to the filtration of financial events. We assume that $\mathbb{T}$ and $\mathbb{G}$ are independent w.r.t the probability measure $\emph{P}$ and that, for every $t$, $\mathcal{F}_t$ is generated by $\mathcal{T}_t$ and $\mathcal{G}_t$, hence $\mathcal{T}_t, \mathcal{G}_t \subset \mathcal{F}_t$ for every $t$.\\
	Therefore, the state price deflator $\bm{\varphi}$ has a product structure
	\begin{equation}
		\varphi_t=\varphi_t^\mathbb{T}\cdot\varphi_t^\mathbb{G}
	\end{equation}
\end{assumption}

\section{The model framework}

\begin{assumption}[\textit{Model assumptions}]
	\label{ModAssump}
We consider now a equity-linked endowment policy with constant annual premiums that guarantee the beneficiary a stochastic amount if the insured dies and a capital if the policyholder survives up to maturity $t=n$. For the sake of simplicity, we assume the same structure of guarantee for both benefits, but the model can be easily extended to different cases.\\
We build a model framework coherent with the market-consistent valuation of Solvency II and we assume :
\begin{enumerate}
	\item In every $t$, with $t\in[0,n-1]$, the undertaking collects the premiums at the beginning of the year according to the cohort who survived at time $t$ and to the sums insured;
	\item Claims that occur in the period $t\in(t,t+1]$ depend on the realizations of i.i.d. Bernoulli random variables and are paid at the end of the one-year period.
\end{enumerate}

\end{assumption}

Considering a generic valuation instant $t$, we present a definition of the future in and out cashflows that is consistent with the aforementioned model assumptions, in particular that premiums are paid at the beginning of each time span and benefits are paid at the end. We denote this quantity with $\textbf{X}_{(t)}=\left[X_\tau\right]_{\tau=t,...,n}$ where each element $X_\tau$ is defined as follows:
\begin{equation}
	X_\tau  =
	\begin{cases}
		-X_\tau^{in} & \text{if $\tau=t$,} \\
		X_\tau^{out}-X_\tau^{in} & \text{if $\tau >t$,}
	\end{cases}
	\label{eq:splitofX}
\end{equation}
where
\begin{equation}
	\label{eq:f12}
	X_\tau^{out}=\langle\left(\textbf{S}_{\tau-1}\right)^\intercal,\bm{\Lambda}^B_{\tau-1},U^{out,\; \tau}_\tau \rangle
\end{equation}
and
\begin{equation}
	\label{xin}
	X_\tau^{in}=\langle\left(\textbf{S}_\tau\right)^\intercal,\bm{\mathds{1}},U^{in,\; \tau}_\tau \rangle
\end{equation}
In formula (\ref{eq:f12}), $\textbf{S}_{\tau-1}$ is a $l_0\times1$ vector of insured sums as specified in Definition \ref{def:coorte}, $\bm{\Lambda}_{\tau-1}^B$ is a $l_0\times1$ $\mathbb{T}$-adapted matrix of insurance technical variables which elements are $\mathbb{I}_{k,\tau-1}^D$ with $k\in[1,l_0]$. Because of Definition \ref{def:coorte}, they are i.i.d. Bernoulli's random variables with parameter equals to $q_{x+\tau-1}$, i.e, the (realistic) death probability of the policyholders. \\
We assume that $\left(U_\tau^{out,\;t}\right)_{t\in[0,n]}$ is a $\mathbb{G}$-adapted stochastic process and represents the process of the financial portfolio $\mathcal{U}^{out}_\tau$ used to replicate the outflow at time $\tau$; moreover, we specify that the superscript $t$ denotes that it is the price of $\mathcal{U}^{out}_\tau$ at time $t$ and, lastly, that the assumption is completely identical for $\left(U_\tau^{in,\;t}\right)_{t\in[0,n]}$. \\
In case $\textbf{S}_{\tau-1}\in\left\{0,1\right\}$, the term $\langle \left(\textbf{S}_{\tau-1}\right)^\intercal,\bm{\Lambda}^B_{\tau-1} \rangle$ represents the r.v. number of deaths in the period $(\tau-1,\tau]$ and it is consistent with the formulation provided in \cite{wuthrich2010market}.\\
As regards to formula (\ref{xin}), $\bm{\mathds{1}}$ is the $l_0\times1$ unitary vector and $U_\tau^{in,\;\tau}$ is a $\mathbb{G}$-adapted stochastic process related to the process of the financial portfolio $\mathcal{U}^{in}_\tau$ used to replicate inflows collected at time $\tau$.\\
As in \cite{wuthrich2010market} and \cite{wuthrich2013financial}, we are assuming a product structure of insurance cash flows $\textbf{X}$, splitting them into a headgeable $\mathcal{G}_t$-measurable part and a non-hedgleable $\mathcal{T}_t$-measurable part.\\
It is therefore possible to extend formula (\ref{generic_reserve}) to the cohort context to calculate the reserve of outstanding premiums and liabilities at a generic time $t$.\\
First of all, let us denote with $\mathbb{I}_{k,t}^L$ a Bernoulli random variable with parameter $p_{x+t}$, equal to the second order annual survival probability of a policyholder aged $x$ at the inception. The r.v. assumes the value 1 if the policyholder $k$ survives; moreover we assume  policyholders are i.i.d.. Consequently we have the r.v. $\mathbb{I}_{k,t}^D=1-\mathbb{I}_{k,t}^L$ for describing the death of the policyholder.

\begin{definition}[\textit{The mathematical reserve for outstanding premiums and liabilities}]
Therefore, we define the mathematical reserves for outstanding premiums and liabilities at time $t$ as
\begin{equation}
		\label{eq:reserve_cohort}
	\begin{aligned}
	\mathcal{R}_t&=\mathcal{Q}_t\left(\textbf{X}_{(t)}\right)\\
	&=\langle \emph{E}\left(\textbf{S}_t\big|\mathcal{T}_t\right)^\intercal,\dfrac{1}{\varphi_t^\mathbb{T}}\emph{E}\left[\left(\bm{\varphi}_{(t+1)}^\mathbb{T}\circ\bm{\Lambda}_{(t)}^L\circ \bm{\Lambda}_{(t)}^B \right)\big|\mathcal{T}_t\right] , \dfrac{1}{\varphi_t^\mathbb{G}}\emph{E}\left[\left(\bm{\varphi}_{(t+1)}^\mathbb{G}\circ\textbf{U}_{(t+1)}^{out}\right)\big|\mathcal{G}_t\right] \rangle \\
	&-\langle \emph{E}\left(\textbf{S}_t\big|\mathcal{T}_t\right)^\intercal, \dfrac{1}{\varphi_{t}^\mathbb{T}}\emph{E}\left[\left(\bm{\varphi}_{(t)}^\mathbb{T}\circ\bm{\Lambda}_{(t)}^L\right)\big|\mathcal{T}_t\right], \dfrac{1}{\varphi_{t}^\mathbb{G}}\emph{E}\left[\left(\bm{\varphi}_{(t)}^\mathbb{G}\circ\textbf{U}_{(t)}^{in}\right)\big|\mathcal{G}_t\right]\rangle
		\end{aligned}
\end{equation}
where $\bm{\Lambda}_{(t)}^L=\left[\lambda^L_{(t),k,\tau}\right]$, $\bm{\Lambda}_{(t)}^B=\left[\lambda^B_{(t),k,\tau}\right]$ with $k=1,...,l_{0}$, $\tau=1,...,n-t$ are $l_0\times (n-t)$ matrices that consider the survival of the policyholder and the unitary benefit in a endowment contract\footnote{It is possible to rewrite $\bm{\Lambda}_{(t)}^B$ for other types of policies. For instance, for a Term Insurance the elements of the last column (i.e. $\tau=n-t$) would be $\mathbb{I}_{k,n-1}^D$. For a Pure Endowment, all the elements of the matrix would be equal to $0$, except those of the last column equal to $\mathbb{I}_{k,n-t}^L$.}, respectively.
We define indeed 
\begin{equation}
	\lambda^L_{(t),k,\tau}  =
	\begin{cases}
		1 & \text{if $\tau=1$,} \\
		\prod_{h=0}^{\tau-2}\mathbb{I}_{k,t+h}^L & \text{if $\tau \geq 1$,}
	\end{cases}
	\label{eq:lambdaL}
\end{equation}

and
\begin{equation}
	\lambda^B_{(t),k,\tau}  =
	\begin{cases}
		\mathbb{I}_{k,t+\tau-1}^D & \text{if $\tau< n-t$,} \\
		1 & \text{if $\tau = n-t $,}
	\end{cases}
	\label{eq:lambdaL}
\end{equation}
Therefore, notice that the Hadamard product $\left(\bm{\Lambda}_{(t)}^L\circ \bm{\Lambda}_{(t)}^B\right)$ results in a $l_0\times (n-t)$ indicator matrix, which (stochastic) elements assume the value $1$ if the generic policyholder is entitled to the benefit at the time specified on the column of the matrix. With a slight abuse of notation, we specify that $\bm{\varphi}^\mathbb{T}_{t+1}$ is also a $l_0\times (n-t)$ matrix, which columns are composed of the same deflators.
\end{definition}
\section{The Cohort Valuation Portfolio}
The purpose of this section is to show the construction of a valuation portfolio (VaPo) and to obtain its value. Given a certain series of future cash flows, a Valuation Portfolio is a portfolio of financial instruments that replicate the specific cash flow. Therefore, we present the steps for building a VaPo which final result is consistent with the cohort approach.
\begin{steps}
	\item \textit{Choice of financial instruments}: The first step concerns the choice of financial basis: with reference to formula (\ref{eq:splitofX}), we are interested in two groups of financial instruments: those for replicating outflows and those for inflows. As regards the former, the choice depends exclusively on the value of the outflows: in the most generic way, here we define with $\mathcal{U}^{out}_t$ the portfolio of financial instruments to replicate the outflow $X_t^{out}$. The $\mathcal{U}^{out}_t$ portfolio can be understood as a linear combination of weights $y^{out}_{i,t}$ of financial instruments available on the market $\mathcal{M}$:
	\begin{equation}
		\label{market_ins}
		\mathcal{U}^{out}_t=\sum_{i\in\mathcal{M}}y_{i,t}^{out}\cdot\mathcal{U}^{out}_{i,t}
	\end{equation}
	In an analogous way, with reference to the inflows, we have:
	\begin{equation}
		\label{market_ins_in}
		\mathcal{U}^{in}_t=\sum_{i\in\mathcal{M}}y_{i,t}^{in}\cdot\mathcal{U}^{in}_{i,t}
	\end{equation}
	\item \textit{Determination of the number of portfolio shares to be held to replicate the cash flows} After defining the financial instruments to replicate outflows and premiums, it is necessary to identify the best estimate of the number of such instruments that the insurance company must hold to replicate the general future cash flows $\textbf{X}_{(t)}$
	In formulas:
	\begin{equation}
		\label{eq:VapoStep2}
		\begin{aligned}
			\textbf{X}_{(t)}\rightarrowtail&VaPo_t\left(\textbf{X}_{(t)}\right)=\\
			&=\langle \emph{E}\left(\textbf{S}_t\big|\mathcal{T}_t\right)^\intercal,\emph{E}\left[\left(\bm{\Lambda}_{(t)}^L\circ \bm{\Lambda}_{(t)}^B \right)\big|\mathcal{T}_t\right] , \bm{\mathcal{U}}_{(t+1)}^{out}\rangle \\
			&-\langle \emph{E}\left(\textbf{S}_t\big|\mathcal{T}_t\right)^\intercal, \emph{E}\left[\left(\bm{\Lambda}_{(t)}^L\right)\big|\mathcal{T}_t\right],\bm{\mathcal{U}}_{(t)}^{in}\rangle
		\end{aligned}
	\end{equation}
\item \textit{The account principle to evaluate financial instruments} This last step presents an accounting principle which associates a monetary value to each financial instrument. Of all the possible accounting principles, we choose to use a fair value valuation consistent with the Solvency II framework, where each financial instrument is valued at the traded market value. We indicate this accounting principles with $\mathcal{E}_t$ with $t\in[0,n]$ and we define the accounting value of the cash-flows as follows:
\begin{equation}
	\label{VapoStep3}
	\begin{aligned}
		\textbf{X}_{(t)}\rightarrowtail&\;\mathcal{Q}_t\left(\textbf{X}_{(t)}\right)=\mathcal{E}_t\left(VaPo_t\left(\textbf{X}_{(t)}\right)\right)\\
		&=\langle \emph{E}\left(\textbf{S}_t\big|\mathcal{T}_t\right)^\intercal,\emph{E}\left[\left(\bm{\Lambda}_{(t)}^L\circ \bm{\Lambda}_{(t)}^B \right)\big|\mathcal{T}_t\right] , \mathcal{E}_t\left(\bm{\mathcal{U}}_{(t+1)}^{out}\right)\rangle \\
		&-\langle \emph{E}\left(\textbf{S}_t\big|\mathcal{T}_t\right)^\intercal, \emph{E}\left[\left(\bm{\Lambda}_{(t)}^L\right)\big|\mathcal{T}_t\right],\mathcal{E}_t\left(\bm{\mathcal{U}}_{(t)}^{in}\right)\rangle
	\end{aligned}
\end{equation}
Introducing the values of financial instruments, we have:
\begin{equation}
		 \mathcal{Q}_t\left(\textbf{X}_{(t)}\right)=\langle\emph{E}\left(\textbf{S}_t\big|\mathcal{T}_t\right)^\intercal,\emph{E}\left[\left(\bm{\Lambda}_{(t)}^L\circ \bm{\Lambda}_{(t)}^B \right)\big|\mathcal{T}_t\right] , \textbf{U}_{(t+1)}^{out,\; t}\rangle 
		-\langle \emph{E}\left(\textbf{S}_t\big|\mathcal{T}_t\right)^\intercal, \emph{E}\left[\left(\bm{\Lambda}_{(t)}^L\right)\big|\mathcal{T}_t\right],\textbf{U}_{(t)}^{in,\;t}\rangle
\end{equation}
\end{steps}
In insurance practice it is very common to use an \textit{ad hoc} demographic table for pricing. This table, called the first order demographic base, leads to a distorted probability (far from any best estimate) whose use allows the creation of an expected demographic profit. Considering a generic indicator $\mathbb{I}_{k,0}^L$ of the matrix $\bm{\Lambda}_{(0)}^L$, we define
\begin{equation}
	(\varphi_1^\mathbb{T}/\varphi_0^\mathbb{T})=\tilde{\varphi}_1^\mathbb{T}
\end{equation}
and
\begin{equation}
	\tilde{\varphi}_1^\mathbb{T}(1)=\dfrac{_1p^*_{x}}{_1p_{x}}
\end{equation}
\begin{equation}
	\tilde{\varphi}_1^\mathbb{T}(0)=\dfrac{1-_1p^*_{x}}{1-_1p_{x}}
\end{equation}	

\noindent As known in the literature (see, e.g,  \cite{olivieri2015introduction}), premiums of a policy are calculated as the expected present value of the benefits, computed with prudent technical bases (also called first-order bases).  \\
In this paper we assume that the insurance company prices the policies of the cohort using a safety loading only with reference to the technical risk, i.e., the demographic one. In other words, the value of financial instruments in the Cohort VaPo protected are calculated throughout the market-consistent accounting principle $\mathcal{E}$. Typically, in the traditional insurance market, insurance companies ensure the non-certainty of default (ruin probability theorem) by inserting implicit safety loading in the pricing phase. For instance, considering a without profit traditional policy it is possible to use deflators of the type
\begin{equation}
	\varphi_t^\mathbb{G}=(1+i^*)^{-t}
\end{equation}
where $i^*$ is a return rate (first order financial basis) that the company is likely to be able to earn on the capital market through the investment of premiums and mathematical reserves. In this paper, we propose a premium calculation structure where the undertaking sells the collateral at exactly the market price. The purpose is to assess the Solvency Capital Requirement linked to demographic risk and the use of implicit safety loading influences the risk profile of undertaking. Our goal is to evaluate only the demographic risk (and therefore the profit), eliminating an additional source of profit of a purely financial nature.\\

\begin{definition}[\textit{Cohort VaPo protected and premiums calculus}]
	\label{def:prem}

The cohort VaPo protected, obtained by replacing the insurance technical risk by their probability distorted conditional expectations at time $0$, is defined as follows:

\begin{equation}
	\label{eq:VapoProt}
	\begin{aligned}
		\textbf{X}_{(0)}\rightarrowtail&VaPo_0^{prot}\left(\textbf{X}_{(0)}\right)=\\
		&=\langle \left(\textbf{s}_0\right)^\intercal,\emph{E}\left[\left(\bm{\varphi}_{(1)}^\mathbb{T}\circ\bm{\Lambda}_{(0)}^L\circ \bm{\Lambda}_{(0)}^B \right)\big|\mathcal{T}_0\right] , \bm{\mathcal{U}}_{(1)}^{out}\rangle \\
		&-\langle \left(\textbf{s}_0\right)^\intercal, \emph{E}\left[\left(\bm{\varphi}_{(0)}^\mathbb{T}\circ\bm{\Lambda}_{(0)}^L\right)\big|\mathcal{T}_0\right],\bm{\mathcal{U}}_{(0)}^{in}\rangle
	\end{aligned}
\end{equation}
\end{definition} 

\noindent Hence, by formula (\ref{market_ins_in}), weights $y_{i,t}^{in}$ are calculated as a solution of the equation
\begin{equation}
	\mathcal{E}_0\left(VaPo_0^{prot}\left(\textbf{X}_{(0)}\right)\right)=0
\end{equation}
or, analogously
\begin{equation}
	\label{eq:VapoProtExtended}
	\langle \left(\textbf{s}_0\right)^\intercal,\emph{E}\left[\left(\bm{\varphi}_{(1)}^\mathbb{T}\circ\bm{\Lambda}_{(0)}^L\circ \bm{\Lambda}_{(0)}^B \right)\big|\mathcal{T}_0\right] , \textbf{{U}}_{(1)}^{out,\;0}\rangle =\langle \left(\textbf{s}_0\right)^\intercal, \emph{E}\left[\left(\bm{\varphi}_{(0)}^\mathbb{T}\circ\bm{\Lambda}_{(0)}^L\right)\big|\mathcal{T}_0\right],\textbf{U}_{(0)}^{in,\;0}\rangle
\end{equation}
We now present a definition of best estimate liabilities that is consistent with both the market-consistent valuation of Solvency II and with Assumptions \ref{ModAssump}.

\begin{definition}[\textit{Best estimate liabilities of a cohort of endowment contracts}]
We define the best estimate of liabilities of a cohort of endowment contracts with an annual premium as follows:
\begin{equation}
	\begin{aligned}
		\mathcal{R}_t=\mathcal{Q}_t\left(\textbf{X}_{(t)}\right)=&\mathcal{E}_t\left(VaPo_t\left(\textbf{X}_{(t)}\right)\right)\\
		&=\langle \emph{E}\left(\textbf{S}_t\big|\mathcal{T}_t\right)^\intercal,\emph{E}\left[\left(\bm{\Lambda}_{(t)}^L\circ \bm{\Lambda}_{(t)}^B \right)\big|\mathcal{T}_t\right] , \mathcal{E}_t\left(\bm{\mathcal{U}}_{(t+1)}^{out}\right)\rangle\\
		&-\langle \emph{E}\left(\textbf{S}_t\big|\mathcal{T}_t\right)^\intercal, \emph{E}\left[\left(\bm{\Lambda}_{(t)}^L\right)\big|\mathcal{T}_t\right],\mathcal{E}_t\left(\bm{\mathcal{U}}_{(t)}^{in}\right)\rangle
	\end{aligned}
	\label{be_liab}
\end{equation}
The case of single premium can be obtained as a specific case of formula (\ref{be_liab}) where second term should be neglected.
\end{definition}
Comparing formulas (\ref{eq:VapoProtExtended}) and (\ref{be_liab}),net of the difference relating to the instant of valuation, it is noticeable the absence of $\bm{\varphi}_{(0)}^\mathbb{T}$ in formula (\ref{be_liab}). The best estimate reserve is indeed defined as the expected present value of the cash flows, using the most up-to-date demographic information and market prices (under the non-arbitrage assumption).

\section{Demographic risk and Solvency}
In order to introduce the proposal for the quantification of the capital requirement related to demographic risk, some distinctive features of the legislation (Solvency II) are presented below, highlighting the aspects that influence the structural characteristics of the proposed model.\\
In particular, within the several sources of risks considered by the Solvency II regulation, we provide here a stochastic model aimed at assessing the capital requirement related to both longevity risk and mortality risk. These risks are defined by evaluating both the risk of loss or of adverse change in the value of insurance liabilities, resulting from changes in the level, trend or volatility of mortality rates (see Art. 105 of \cite{solvency2}). Consistently with the legislation, a model is therefore presented that focuses on the two risks of losses and unfavourable variations of insurance liabilities and that considers both the effects of longevity and mortality.

\subsection{The model framework}
\begin{figure}[!ht]
	\centering
	\includegraphics[width=140mm]{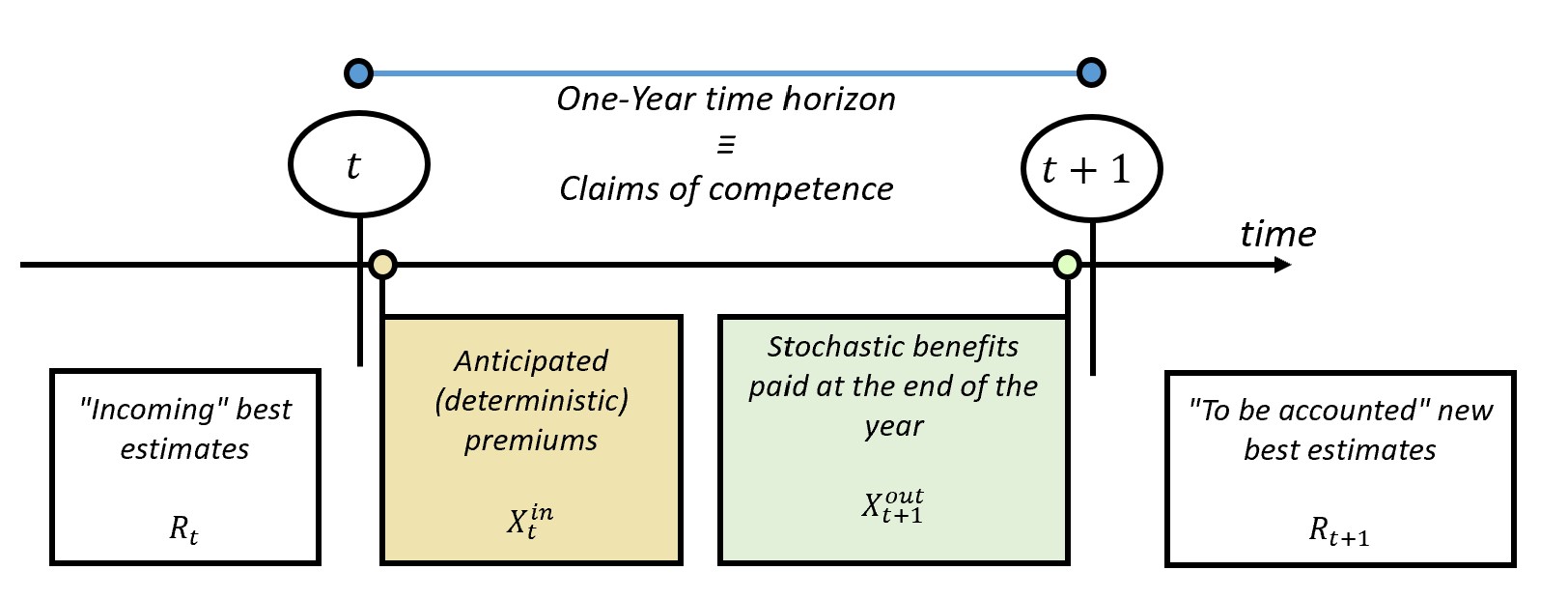}
	\caption{A graphical representation of the book entries in a generic year $ t $ and considering an annual time horizon}
	\label{Fig:bil}
\end{figure}

Examining a generic period $(t,t+1]$, as shown in Figure \ref{Fig:bil}, we consider to have available, at the beginning of the year, the best estimate $\mathcal{R}_t$ (see (\ref{be_liab})) and the deterministic premiums $X_t^{in}$. These amounts, properly invested, should cover the r.v. claims $X_{t+1}^{out}$ and the new reserves $\mathcal{R}_{t+1}$.
\\The starting point therefore coincides considering the quantity $V_{t+1}$ defined as
\begin{equation}
	\label{vtplus1}
	\begin{aligned}
	V_{t+1}=&
		\langle \emph{E}\left(\textbf{S}_t\big|\mathcal{T}_t\right)^\intercal,\emph{E}\left[\left(\bm{\Lambda}_{(t)}^L\circ \bm{\Lambda}_{(t)}^B \right)\big|\mathcal{T}_t\right] , \textbf{U}_{(t+1)}^{out,\; t+1}\rangle\\ 
		&-\langle \emph{E}\left(\textbf{S}_t\big|\mathcal{T}_t\right)^\intercal, \emph{E}\left[\left(\bm{\Lambda}_{(t+1)}^L\right)\big|\mathcal{T}_t\right],\textbf{U}_{(t+1)}^{in,\;t+1}\rangle
			\end{aligned}
\end{equation}
Therefore, we assume to buy $VaPo_t\left(\textbf{X}_{(t)}\right)$ at time $t$ at its market price $\mathcal{R}_t$. In addition to this quantity, we collect the inflow $X_t^{in}$ from the policyholders. At the end of the time span, i.e. in $t+1$, the market price total value is $V_{t+1}$.\\
We define the Claims Development Result (CDR) $CDR_{t+1}$ as 
\begin{equation}
	\label{CDR}
	CDR_{t+1}=V_{t+1}-X_{t+1}^{out}-\mathcal{R}_{t+1}
\end{equation}
We observe that in formula (\ref{CDR}) the available financial information is described by $\mathcal{G}_{t+1}$, while the insurance technical one is represented by $\mathcal{T}_{t+1}$. Moreover, within $V_{t+1}$ the expected value is calculated on the basis of $\mathcal{T}_t$, while in $\mathcal{R}_{t+1}$ the expected value of the technical risk is conditioned on the information contained in $\mathcal{T}_{t+1}$.\\
We then introduce the quantity $\hat{\mathcal{R}}_{t+1}$
\begin{equation}
	\begin{aligned}
	\hat{\mathcal{R}}_{t+1}=&\langle \emph{E}\left(\textbf{S}_{t+1}\big|\mathcal{T}_{t+1}\right)^\intercal,\emph{E}\left[\left(\bm{\Lambda}_{(t+1)}^L\circ \bm{\Lambda}_{(t+1)}^B \right)\big|\mathcal{T}_{t}\right] , \mathcal{E}_{t+1}\left(\bm{\mathcal{U}}_{(t+2)}^{out}\right)\rangle\\
	&-\langle \emph{E}\left(\textbf{S}_{t+1}\big|\mathcal{T}_{t+1}\right)^\intercal, \emph{E}\left[\left(\bm{\Lambda}_{(t+1)}^L\right)\big|\mathcal{T}_{t}\right],\mathcal{E}_{t+1}\left(\bm{\mathcal{U}}_{(t+1)}^{in}\right)\rangle
	\end{aligned}
\end{equation}
that represents the best estimate in $t+1$ under the assumption that the technical risk is conditioned on the information contained in $\mathcal{T}_t$.\\
Hence, we split the CDR in two components as follows:
\begin{equation}
	\label{CDR_idios}
	CDR_{t+1}^{Idios}=V_{t+1}-X_{t+1}^{out}-\hat{\mathcal{R}}_{t+1}
\end{equation}
and 
\begin{equation}
	\label{CDR_trend}
	CDR_{t+1}^{Trend}=\hat{\mathcal{R}}_{t+1}-\mathcal{R}_{t+1}
\end{equation}
It could be noticed that two sources of uncertainties of demographic nature are present. The first is given by the accidental mortality of the cohort, related the volatility due to the mortality/longevity probabilities (i.e, the idiosyncratic risk). The second source is related to the fact that new information available during the year could lead to a modification of the best estimate at the end of the year. For instance, this aspect occurs when a trend in mortality or longevity rates is observed. This risk will be referred to as trend risk.

\subsection{The idiosyncratic risk}
We focus here on the idiosyncratic risk defined as the risk of monetary loss resulting from changes in level or volatility of mortality rates. It concerns both polices whose benefit is linked to the death of the policyholder and policies whose benefit is linked to the survival of the policyholder.
\begin{definition}
\label{def:idios}
In a one-year time horizon, we define the idiosyncratic profit and loss extending formula (\ref{CDR_idios}):
	\begin{equation}
		\begin{aligned}
			CDR_{t+1}^{Idios}= &  \langle \emph{E}\left(\textbf{S}_{t}\big|\mathcal{T}_{t}\right)^\intercal,\emph{E}\left[\left(\bm{\Lambda}_{(t)}^L\circ \bm{\Lambda}_{(t)}^B \right)\big|\mathcal{T}_t\right] , \textbf{U}_{(t+1)}^{out,\; t+1}\rangle \\
			&-\langle \emph{E}\left(\textbf{S}_{t}\big|\mathcal{T}_{t}\right)^\intercal, \emph{E}\left[\left(\bm{\Lambda}_{(t+1)}^L\right)\big|\mathcal{T}_t\right],\textbf{U}_{(t+1)}^{in,\;t+1}\rangle \\
			& -\langle\left(\textbf{S}_{t}\right)^\intercal,\bm{\Lambda}^B_{t},U^{out,\; t+1}_{t+1} \rangle \\
			&-\langle \emph{E}\left(\textbf{S}_{t+1}\big|\mathcal{T}_{t+1}\right)^\intercal,\emph{E}\left[\left(\bm{\Lambda}_{(t+1)}^L\circ \bm{\Lambda}_{(t+1)}^B \right)\big|\mathcal{T}_t\right] , \textbf{U}_{(t+2)}^{out,\;t+1}\rangle\\
			&+\langle \emph{E}\left(\textbf{S}_{t+1}\big|\mathcal{T}_{t+1}\right)^\intercal, \emph{E}\left[\left(\bm{\Lambda}_{(t+1)}^L\right)\big|\mathcal{T}_t\right],\textbf{U}_{(t+1)}^{in,\;t+1}\rangle
		\end{aligned}
		\label{idios}
	\end{equation}
As usual in a life insurance balance-sheet (see, e.g.,  \cite{wuthrich2010market} and \cite{wuthrich2013financial}) the profit is here defined as the sufficiency of initial best estimate and premiums collected to cover the payment of benefits, due to the claims occurred during the year, and the assessment of the final best estimate.  Notice that in the idiosyncratic component we want to catch only the risk due to the volatility of deaths in the year $(t,t+1]$, that affects both the payments in $t+1$ and the sums insured used in the new best estimates calculation. 
Therefore, we assume that the insurance company does not change its assumptions on future mortality, considering any change in mortality during the year to be accidental. Indeed, the effect on the final best estimate due to a revision of demographic assumptions is devoted to the trend risk.\\
We provide in the following Theorem a compact formulation of the r.v. idiosyncratic risk. Proof is reported in \ref{sec:apA}.
	\begin{theorem}[\textit{Compact formulation of Idiosyncratic Risk}]
		\label{Teorema_1}
		Considering a generic cohort of policyholders consistent with Definition \ref{def:coorte}, if the premiums are paid in advance, the claims are paid at the end of the occurrence year and the best estimates at the end of the year are calculated on the same demographic and financial bases used at the beginning year, it is possible to define the idiosyncratic risk as
		\begin{equation}
		CDR_{t+1}^{Idios}=\langle\left[ \emph{E}\left(\textbf{S}_t\circ  \bm{\Lambda}_{t}^B \big|\mathcal{T}_t\right)-\emph{E}\left(\textbf{S}_t\circ  \bm{\Lambda}_{t}^B\big|\mathcal{T}_{t+1}\right)\right]^\intercal,\mathds{1} \cdot \eta_{t+1} \rangle
			\label{eq:Idios_compact}
		\end{equation}
where  $ \eta_{t+1} =\left( U_{t+1}^{out,\;t+1}-\beta_{t+1}\right)$ with $\beta_{t+1}$ best estimate rate of a policyholder of the cohort calculated in $t+1$ with technical bases estimated in $t$; or, analogously, $\beta_{t+1}$ is the generic element of $\bm{\beta}_{t+1} = \langle \emph{E}\left(\bm{\Lambda}_{(t+1)}^L\circ\bm{\Lambda}_{(t+1)}^B\big|\mathcal{T}_t\right), \textbf{U}_{(t+2)}^{out,\;t+1}\rangle-\langle \emph{E}\left(\bm{\Lambda}_{(t+1)}^L\big|\mathcal{T}_t\right),\textbf{U}_{(t+1)}^{in,\;t+1}\rangle $.
	\end{theorem}
	According to formula (\ref{eq:Idios_compact}), if the accidental (idiosyncratic) mortality of the year is higher than expected, the insurer will have to cope with a greater amount of claims $U_{t+1}^{out,\;t+1}$. In this scenario, the firm will release best estimate liabilities which unit rate for each policyholder is $\beta_{t+1}$.\\
	It is noteworthy that we can interpret the term $\eta_{t+1}$  as the expected value at time $t$ of the Sum-at-Risk rate at time $t+1$ of a policyholder. Notice that this result generalizes the simplified case of non equity-linked contracts where $U_{t+1}^{out,\;t+1}$ would be equal to one (see, e.g. \cite{savelli2013risk}, \cite{clemente2021bridge} and \cite{clemente2022stochastic}).\\
	We can focus now on the main cumulants of $CDR_{t+1}^{Idios}$. In particular, using the tower property:
	\begin{equation}
		\emph{E}\left(CDR_{t+1}^{Idios}\big|\mathcal{F}_t\right)=0
		\label{eq:ev_idios}
	\end{equation}
	Therefore, on average the best estimate at time $ t $ added to the premiums, is sufficient to meet the payments of the expected benefits and the provision of new best estimates calculated on the same demographic basis. Figure \ref{Fig:cdr} shows that, if the accidental mortality of the year is different\footnote{The sign of the variation in mortality that produces a loss depends on the type of contract} from the expected one, a profit or loss is created.
	\begin{figure}[ht!]
		\centering
		\includegraphics[width=120mm]{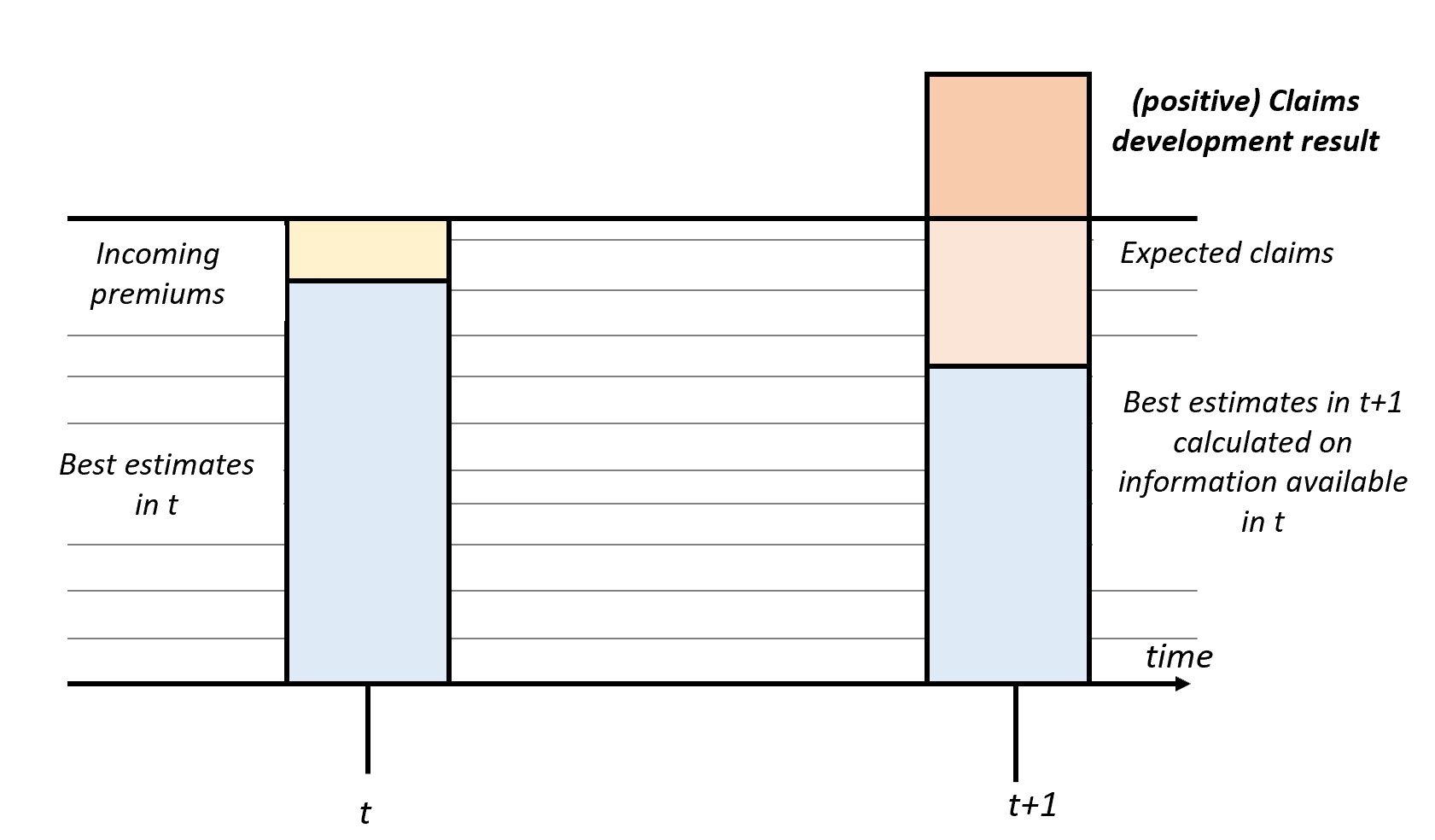}
		\caption{Claims development result representation in $[t,t+1]$ }
		\label{Fig:cdr}
	\end{figure}
\end{definition}
\noindent In this way, the idiosyncratic profit definition presented in formula (\ref{idios}) is perfectly consistent with the concept of Claims Development Result, and the result is consistent with \cite{wuthrich2013financial} and \cite{wuthrich2008stochastic}.\\
Considering the random variable defined in formula (\ref{idios}), we prove (see \ref{sec:apA} for details) that the standard deviation of the idiosyncratic profit is equal to 
\begin{equation}
	Var\left(CDR_{t+1}^{Idios}\big|\mathcal{F}_t\right) =\left( l_{t}\cdot\ q_{x+t}\cdot(1-q_{x+t})\cdot \bar{\textbf{S}}_t^2\right)\cdot \emph{E}\left(\eta^2_{t+1}\big|\mathcal{G}_t\right)
	\label{eq:sigma_y}
\end{equation}
where  $\bar{\textbf{S}}_t^j$ denotes the $j$-raw moment of the $\mathcal{F}_t$-adaptable vector $\textbf{S}_t$.
We therefore observe that the volatility of the idiosyncratic risk increases with the ageing of the cohort\footnote{Only for cohorts with $q_{x+t}<0.5$, therefore all ages except for the extreme ages}, the size of the cohort $l_t$ and a higher variability within the sums insured. Additionally, it depends on the path of the expected Sum-at-Risk rate, that, as well-known, is related to the the type of policies considered. For instance, it has a decreasing behaviour for a endowment contract and a convex one for a term insurance with annual premiums. 
\begin{definition}[\textit{Risk measure}]\label{def:risk}
	We define the generic risk measure (see \cite{artzner1999coherent}) as a function $\varsigma$
	\begin{equation}
		\varsigma:\emph{L}^2(\emph{P}) \rightarrow \mathbb{R}^+; \qquad X \rightarrow \varsigma(X)
	\end{equation}
	General pleasant properties for a risk measure are normalization ($\varsigma(0)=0$), translation invariance ($\varsigma(c+X)=c+\varsigma(X)$), positive homogeneity(if $c>0$, then $\varsigma(c\cdot X)=c\cdot \varsigma(X)$), subadditivity ($\varsigma(X+Y)\leq\varsigma(X)+\varsigma(Y)$), monotonicity (if $X\leq Y$ $\emph{P}$-almost surely, then $\varsigma(X)\leq\varsigma(Y)$), expectation boundedness ($\varsigma(X)>\emph{E}^\emph{P}(-X)$).
\end{definition}
\noindent Although Value at Risk does not satisfy all the properties in Definition \ref{def:risk}, following Solvency II regulation (see Art. 102), the Solvency Capital Requirement can be computed with a partial internal model
as follows:
\begin{equation}
	\varsigma\left(CDR_{t+1}^{Idios}\right)=-VaR_{0.5\%}\left(CDR_{t+1}^{Idios}\right)
\end{equation}
where $VaR_{0.5\%}$ stands for Value at Risk calculated with a confidence level equal to $99.5\%$.\\
Our aim is to also provide a possible undertaking specific approach for quantifying the capital requirement due to idiosyncratic risk. To this end, adapting the formula provided for non-life premium and reserve risk, we define a factor-based structure for capital assessment. Knowing the characteristics of the distribution of $CDR_{t+1}^{Idios}$, it is therefore possible to write
\begin{equation}
	SCR_{L,Idios}=f\left[CDR_{t+1}^{Idios}\right]\cdot \sigma\left(CDR_{t+1}^{Idios}\right) 
	\label{eq:USP_Idios}
\end{equation}
where $f\left[CDR_{t+1}^{Idios}\right]$ is a multiplier, appropriately calibrated, of the standard deviation, calculated as a function of the distribution characteristics of the r.v. idiosyncratic profit.
\subsection{The trend risk}
\noindent Comparing the definitions of the whole demographic profit (see Figure \ref{Fig:bil} and the idyosincratic profit (see formula (\ref{idios})), we have to consider the possibility that the best estimate a time $ t + 1 $, given by $\mathcal{R}_{t+1}$, can be could differ from the reserve accounted in the idiosyncratic profit. \\
\begin{definition}
	We define the random variable demographic profit (or loss) due to trend risk $CDR_{t+1}^{Trend}$ as
	
	\begin{equation}
		\begin{aligned}
			CDR_{t+1}^{Trend} = & \langle\emph{E}\left(\textbf{S}_{t+1}\big|\mathcal{T}_{t+1}\right)^\intercal,  \langle \emph{E}\left(\bm{\Lambda}_{(t+1)}^L\circ\bm{\Lambda}_{(t+1)}^B\big|\mathcal{T}_{t+1}\right)-\emph{E}\left(\bm{\Lambda}_{(t+1)}^L\circ\bm{\Lambda}_{(t+1)}^B\big|\mathcal{T}_{t}\right), \textbf{U}_{(t+2)}^{out,\;t+1}\rangle\\
			&-\langle\emph{E}\left(\bm{\Lambda}_{(t+1)}^L\big|\mathcal{T}_{t+1}\right)-\emph{E}\left(\bm{\Lambda}_{(t+1)}^L\big|\mathcal{T}_{t}\right),\textbf{U}_{(t+1)}^{in,\;t+1}\rangle \rangle 
		\end{aligned}
		\label{eq:trend}
	\end{equation}
\end{definition}
\noindent It is noteworthy that, the Claim Development Result linked to the trend risk depends on two sources of uncertainty: since in $t+1$ the technical information $\mathcal{T}_{t+1}$ is different from that of the previous instant $\mathcal{T}_{t}$, the insurance company can update (albeit to a limited extent) future expectations. Secondly, it is specified that, as in idiosyncratic risk context, the adjustment of technical expectations implies purchases or sales of shares in replicating portfolios: for this reason, trend risk also incorporates a share of financial risk.\\
In this context, the distribution of $CDR_{t+1}^{Trend}$ is therefore influenced by both the accidental mortality of the year (given by $\textbf{S}_{t+1}$) and the revision of the demographic basis.
In conclusion, we highlight that since $\mathcal{F}_t\subset\mathcal{F}_{t+1}$, the tower property implies that
\begin{equation}
	\emph{E}\left(CDR_{t+1}^{Trend}\big|\mathcal{F}_t\right)=0
\end{equation}
\begin{definition} (\textit{SCR for Trend risk})
We define the Solvency Capital requirement for Trend risk as 
	\begin{equation}
		SCR_{L,Trend}=\varsigma\left(CDR_{t+1}^{Trend}\right)
		\label{SCR_trend_finale}
	\end{equation}
Quantile of $CDR_{t+1}^{Trend}$ can be carried out through different methodologies. We propose a simple approach based on Monte Carlo, capable of capturing the aforementioned dependencies between accidental mortality and trend risk.\\
	We propose the following algorithm:
	\begin{enumerate}
		\item Fit a projection model (see, e.g, \cite{lee2000lee}, \cite{brouhns2002poisson}, \cite{cairns2006two}, \cite{renshaw2006cohort}, \cite{plat2011one}, etc.) to forecast expected mortality rates using  train data available at time $t$;
		\item Calculate $\textbf{U}_{(t+2)}^{out,\;t+1}$ and $\textbf{U}_{(t+1)}^{in,\;t+1}$ simulating the underlying financial variables $\textbf{Z}_{t+1}$;
		\item For each policyholder $k$, simulate $ H $ times the r.v. $\mathbb{I}_{k,t}^L$ from i.i.d. Bernoulli distributions with parameter $p_{x+t}$ given by the expected mortality rates obtained at step 1;
		\item Compute for each simulation the stochastic vector $\textbf{S}_{t+1}$ of the sums insured at time $t+1$;
		\item For each simulation, fit the same mortality model used at step 3 on a train set enriched with additional information simulated under real-word probabilities;
		\item Using values at steps 2, 4 and 5, compute for each simulation $CDR_{t+1}^{h,Trend}$ with $h\in[1,H]$;
		\item Calculate formula (\ref{SCR_trend_finale}) as
		\begin{equation}
			SCR_{L,Trend}=-inf\left[CDR_{t+1}^{h,Trend}: \emph{F}_{{CDR}_{t+1}^{Trend}}\left(CDR_{t+1}^{h,Trend}\right)>0.5\%\right]
		\end{equation}
	\end{enumerate} 
	Therefore, having obtained the distribution of $CDR_{t+1}^{Trend}$, through the $H$ simulations, we calculate the quantity $-VaR_{p=0.5\%}\left(CDR_{t+1}^{n,Trend}\right)$  as the opposite of the lower extreme of the obtained vector, weighted by the probabilities of both $\textbf{S}_{t+1}$ and $\textbf{Z}_{t+1}$.
\end{definition}

\section{The case study}
We develop here a case study in order to investigate the behaviour of the model in describing both idiosyncratic and trend risk. The characteristics of the cohort and of the contracts and some preliminary aspects are outlined in Section \ref{prelsub}. Main results regarding idiosyncratic and trend risks are provides in Section \ref{idiores} and \ref{treres}, respectively.

\subsection{Preliminaries}\label{prelsub}
We summarize in Table 1 main characteristics of the cohort analysed. In particular, we consider a cohort of 10,000 policyholders who have subscribed equity-linked endowment policies, each with its own specific sum insured. If the generic $k$-th policyholder dies in the time span $(t-1;t]$ the beneficiary is entitled to receive a benefit equal to $S_{k,t}\cdot max\left(\dfrac{U_t}{U_0},(1+i_{gar})^t\right)$ where $U_t$ is the market value of the underlying asset $\mathcal{U}$ (the equity) at time $t$ and $i_{gar}$ is the minimum guaranteed rate.\\

\begin{table}[!h]
	\label{tab:intro}
	\centering
	\caption{Model parameters}
	\begin{tabular}{|l|c|}
		\hline
		\textbf{Characteristics}            & \textbf{Value}                                     \\
		\hline
		Number of policyholders & 10,000                                     \\ \hline
		Cohort age                 & 50                                        \\ \hline
		Policies duration                 & 10 years                                        \\ \hline
		2nd order demographic base & Lee-Carter applied on 1852-2019 Italy data \\ \hline
		1st order demographic base & 2nd order $q_x$ stressed of 20\%          \\ \hline
		Risk-free rate             & Costant and equals to 2\%                 \\ \hline
		Guaranteed rate $i_{gar}$             & 1\%                 \\ 
		\hline
		Average sum insured             & 100,000.00                 \\ 
		\hline
		CoV of $s_0$             & 2                \\ 
		\hline
	\end{tabular}
\end{table}

Assuming $U_0=1$, the general payoff $max\left(\dfrac{U_t}{U_0},(1+i_{gar})^t\right)$ can be hedged through the ownership of the equity $\mathcal{U}$ and a put option $\mathcal{P}\left(\mathcal{U}, (1+i_{gar})^t\right)$ written on $\mathcal{U}$, with a strike price and a maturity equals to $(1+i_{gar})^t$ and $t$, respectively.\\
According to the number of equities that the company must hold to form the replicating portfolio, we can define the following relation:
\begin{equation}
	\sum_{t=1}^{n-1}\emph{E}\left(D_{x+t}\big|\mathcal{T}_0\right)+\emph{E}\left(L_{x+n}\big|\mathcal{T}_0\right)=l_0
\end{equation}
where $D_{x+t}$ denotes the r.v. number of deaths within the cohort between $t-1$ and $t$ and $L_{x+n}$ the r.v. number of policyholders survived at maturity. \\
\noindent Therefore, $s_{k,0}$ indicates the number of replicating portfolio shares due to the $k$-th policyholder. Hence, the insurance company, by buying exactly $s_{0}=\sum_{k=1}^{l_0}s_{k,0}$ equities, will certainly be able to meet part of the liabilities; the risk of incurring losses for the company concerns only the part of outflows covered by Put options. We represent the whole Hedging Portfolio in Table 2.

\begin{table}[]
	\centering
	\label{tab:HedPort}
	\caption{The Hedging Portfolio bought in $t=0$}
	\begin{tabular}{|l|c|}
		\hline
		\textbf{Financial Instrument }        & \textbf{Amount in the hedging portfolio}        \\ \hline
		$\mathcal{U}$               & $s_0$    \\ \hline
		$\mathcal{P}\left(\mathcal{U}, (1+i_{gar})^1\right)$ & $s_0\cdot$\emph{E}$\left(D_{50+1}\big|\mathcal{T}_0\right)$ \\ \hline
		...           & ...     \\ \hline
		$\mathcal{P}\left(\mathcal{U}, (1+i_{gar})^{10}\right)$ & $s_0\cdot\left[\emph{E}\left(D_{50+10}\big|\mathcal{T}_0\right)+\emph{E}\left(L_{50+10}\big|\mathcal{T}_0\right)\right]$ \\ \hline
	\end{tabular}
\end{table}

 \noindent By assuming a constant risk-free flat rate over time, the prices of Zero Coupon Bonds can be easily determined. However, it is important to note that while the model presented in this paper is applicable to a wide range of inflows, equity-linked policies in insurance practice are typically financed either through a Single Premium or Recurrent Premiums. The utilization of Regular Annual Premiums would necessitate spreading the purchase cost of equities throughout the entire premium duration. In such a scenario, the insurance company would incur a substantial initial outflow before recovering the costs, which is unsustainable given the negative best estimate of the mathematical reserve. This situation could create an incentive for the policyholder to terminate the contract.\\
Furthermore, we assume that we are in an efficient market context, all relevant information is reflected in the prices of the underlying assets, there are neither dividends nor transition costs and prices are described by Geometric Brownian Motions (GBM). In particular 
\begin{equation}
	dU_t=\mu\;U_t\;dt+\sigma\;U_t\;dW_t
\end{equation}
where $W_t$ is a Wiener process. We calibrated the parameters of the GBM in order to obtain
\begin{equation}
	 \begin{cases}
		\emph{E}\left(U_t\right)=(1+15\%)^t &\\
		CoV\left(U_t\right)=1\\
	\end{cases}       
\end{equation}
it was therefore possible to calculate the prices of put options at time $t$ with maturity $m$ with the well-known Black \& Scholes formula.

\subsection{Idiosyncratic risk results}\label{idiores}
In Table 3 we present the results of the simulation model aimed at obtaining the distribution of $CDR_{t+1}^{Idios}$ at different time instants (displayed in Figure 3). First of all, we specify that we performed 1 million simulations for each instant of time, obtaining consistent results. The number of simulations has been indeed validated by verifying that the simulated characteristics of the distribution converge to the exact ones.

\begin{table}[!h]
	\centering
	\label{tab:idiosResults}
		\caption{Numerical results of $CDR_{t+1}^{Idios}$ }
	\begin{tabular}{|l|c|c|c|c|}
		\hline
		$\bm{CDR_{t+1}^{Idios}}$          & \textbf{t=1}       & \textbf{t=3}       & \textbf{t=5}      & \textbf{t=7}       \\ \hline
		Simulated mean               & -1,106    & 700       & -983      & -591      \\ \hline
		Simulated standard deviation & 1,029,540 & 1,197,463 & 1,394,818 & 1,604,043 \\ \hline
		Simulated skewness           & 12.63     & 11,36     & 11,36     & 11,47     \\ \hline
		Solvency Capital Requirement & 3,526,052 & 4,085,160 & 4,845,989 & 5,638,829 \\ \hline
	\end{tabular}
\end{table}
As shown before, the average of the distributions is centered on zero. With reference to the standard deviation, we notice two contrasting effects.
On the one hand, a higher volatility due to an increased death probability over time. On the other hand, the expected sum at risk rate decreases over time. However results show a prevalence of the mortality rates leading to an increase of the standard deviation over time. \\
It is also observed a relevant positive skewness quite stable over time. As a consequence we have an increased capital that is equal to more than 3 times the volatility.

\begin{figure}[ht!]
	\centering
	\includegraphics[width=120mm]{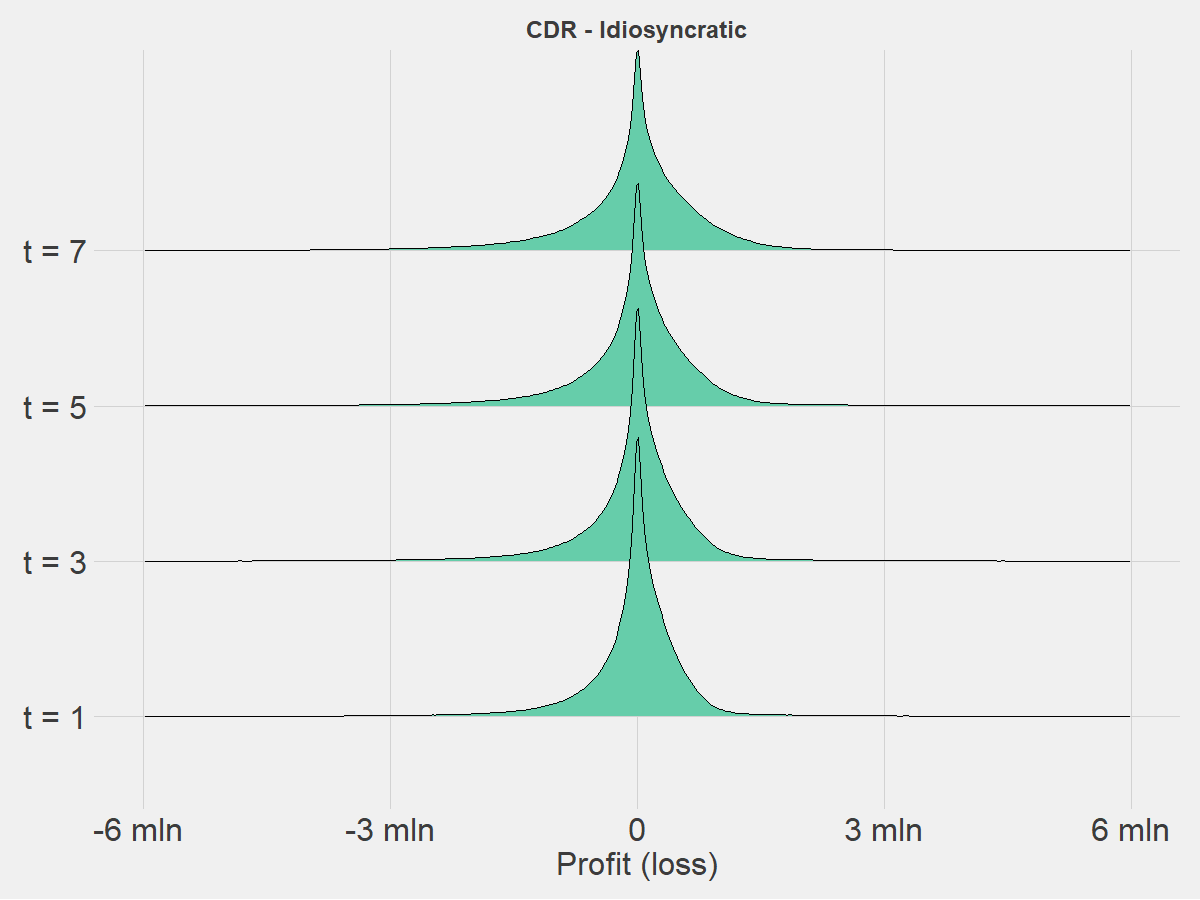}
	\caption{Idiosyncratic Claims Development Result representation in $[t,t+1]$ }
	\label{Fig:cdr}
\end{figure}

\noindent It is noteworthy that formula (\ref{eq:USP_Idios}) shows some common features with the factor-based approach provided for the assessment of idiosyncratic component of longevity risk in Quantitative Impact Study n.2 (see \cite{qis2}). The SCR for idiosyncratic mortality risk was indeed defined in QIS2 as follows:
\begin{equation}
	SCR^{Id,QIS2}=2.58\cdot \sqrt{\dfrac{q\cdot(1-q)}{l}}\cdot Sum-at-Risk.
	\label{QIS2}
\end{equation}
The simulation framework based on the cohort approach therefore highlights two innovative aspects:
\begin{itemize}
 \item The variability of the vector of the insured sums might be considered;
 \item The fact that the distributions of the Claims Development Results for Idiosyncratic Risk have positive skewness and are all leptokurtic (see Figure \ref{Fig:cdr}), imposes the use of a multiplier greater than the 2.58 obtainable under the assumptions of Normal distribution. A better proxy could be the multiplier equal to 3 used for the Premium \& Reserve Risk of the Non-Life Underwriting Risk macro-module.
\end{itemize}

\subsection{Trend Risk results}\label{treres}
	\begin{figure}[ht!]
	\centering
	\includegraphics[width=120mm]{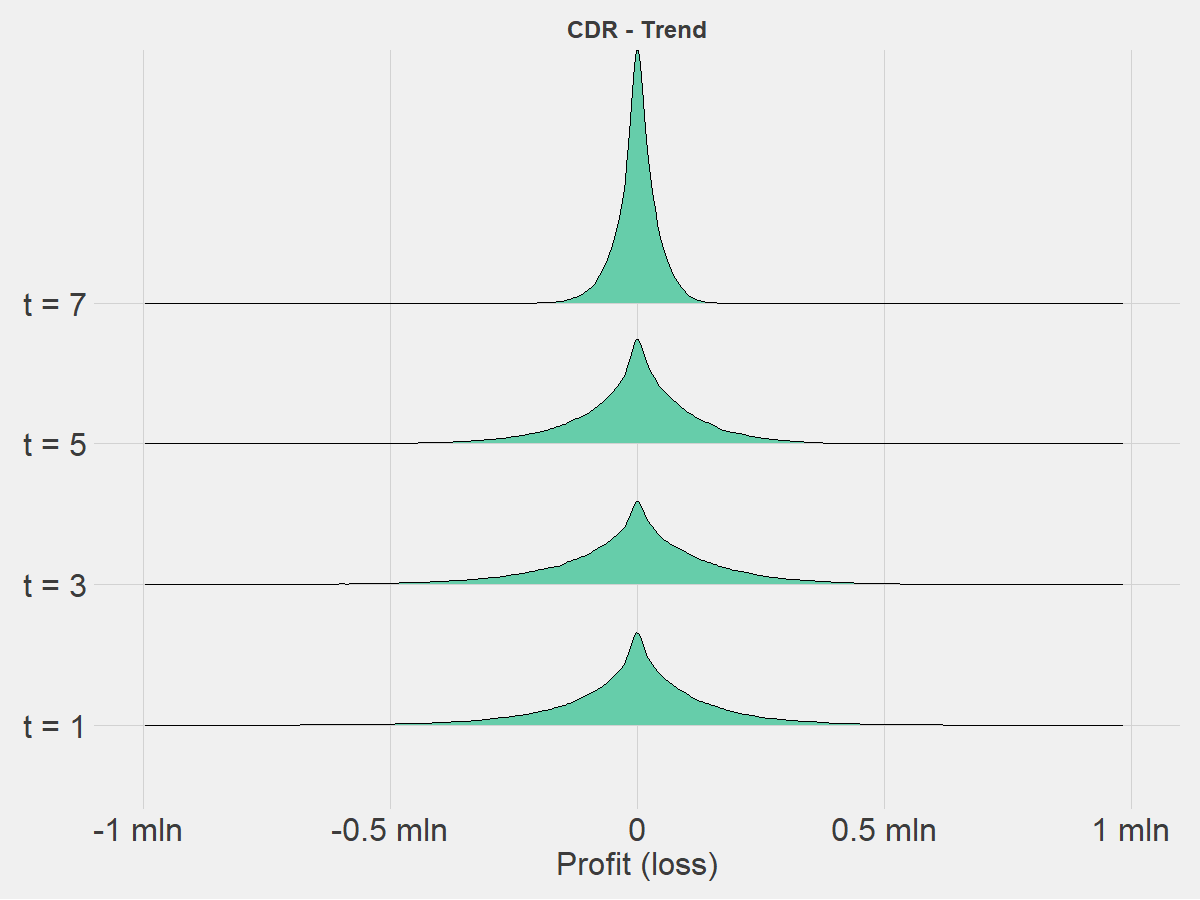}
	\caption{Trend claims development result representation in $[t,t+1]$ }
	\label{Fig:cdr_trend}
\end{figure}
Figure 4 and Table 4 show the results of the simulation model aimed at quantifying the Claims Development Result for Trend Risk. For th sake of simplicity, future mortality rates have been obtained using the Lee-Carter model. This choice is related to the fact that at each simulation it is necessary to re-fit the mortality projection model and, therefore, the computation times are particularly more consistent, despite the matrix representation of the components. However, the approach can be tested with alternative models for forecasting mortality.\\
The results of the model highlight two phenomena: on the one hand, the volatility linked to the new reserve estimates is significantly lower than the idiosyncratic volatility. This phenomenon was foreseeable, since the sample used to estimate the initial table were based on data from 1872 to 2019; an additional unit number/information content can only have a limited effect on updating expectations. On the second hand, we observe that the standard deviation of the CDR for Trend Risk decreases over time. Indeed, any revision of expectations has an increasingly limited effect as the expiry of the policies approaches because a change of expectations affects only a few cash flows.\\ 
\noindent In conclusion, we observe that also for the CDR Trend, 3 times the standard deviation is a good proxy of the Solvency Capital Requirement.
\begin{table}[]
	\centering
	\label{tab:trendResults}
	\caption{Numerical results of $CDR_{t+1}^{Trend}$ }
	\begin{tabular}{|l|c|c|c|c|}
		\hline
		$\bm{CDR_{t+1}^{Trend}}$          & \textbf{t=1}       & \textbf{t=3}       & \textbf{t=5}      & \textbf{t=7}       \\ \hline
		Simulated mean               & -2,269    & -3,374       & -2,948      & 1,772      \\ \hline
		Simulated standard deviation & 158,451 & 149,561 & 125,348 & 41,175 \\ \hline
		Simulated skewness           & -0.31     & -0.32     & -0.33    & -0.34     \\ \hline
		Solvency Capital Requirement & 546,330 & 521,549 & 424,445 & 137,818 \\ \hline
	\end{tabular}
\end{table}

\subsection{Conclusions}
In conclusion, the proposed model incorporates the market-consistent actuarial valuation proposed by Solvency II and effectively computes the key characteristics of both components of demographic risk: idiosyncratic and trend. The model shows an excellent performance when applied to a single cohort of policyholders. Additionally, it can be extended to assess the riskiness of portfolios composed of multiple cohorts, serving as a valuable model point.\\
One notable advantage of the model is its efficient computational framework, leveraging matrix notation to significantly reduce computation times. By utilizing this approach, actuarial calculations can be performed more swiftly, facilitating timely analysis and decision-making processes.\\
Furthermore, we emphasize the importance of considering the volatility of sums insured, element that is neglected by approaches based on classical models for forecasting mortality. This volatility fluctuations has indeed a substantial impact on both the demographic and trend distributions and the inclusions of this aspect in the model allows a more accurate assessment of the associated risks.\\
The numerical analyses show the applicability and effectiveness of the model within the context of Solvency II. It offers valuable insights and practical tools for actuarial professionals to enhance their risk assessment and management capabilities.

\noindent 
\newpage
\bibliographystyle{acm}
\bibliography{BiblThird}

\begin{thebibliography}{10}

\bibitem{artzner1999coherent}
{\sc Artzner, P., Delbaen, F., Eber, J.-M., and Heath, D.}
\newblock Coherent measures of risk.
\newblock {\em Mathematical finance 9}, 3 (1999), 203--228.

\bibitem{BArigou}
{\sc Barigou, K., Bignozzi, V., and Tsanakas, A.}
\newblock Insurance valuation: A two-step generalised regression approach.
\newblock {\em ASTIN Bulletin 52\/} (12 2021), 1--35.

\bibitem{BERNARD2005499}
{\sc Bernard, C., {Le Courtois}, O., and Quittard-Pinon, F.}
\newblock Market value of life insurance contracts under stochastic interest
  rates and default risk.
\newblock {\em Insurance: Mathematics and Economics 36}, 3 (2005), 499--516.

\bibitem{boonen2017solvency}
{\sc Boonen, T.~J.}
\newblock Solvency ii solvency capital requirement for life insurance companies
  based on expected shortfall.
\newblock {\em European actuarial journal 7}, 2 (2017), 405--434.

\bibitem{brennan1976pricing}
{\sc Brennan, M.~J., and Schwartz, E.~S.}
\newblock The pricing of equity-linked life insurance policies with an asset
  value guarantee.
\newblock {\em Journal of Financial Economics 3}, 3 (1976), 195--213.

\bibitem{brouhns2002poisson}
{\sc Brouhns, N., Denuit, M., and Vermunt, J.~K.}
\newblock A poisson log-bilinear regression approach to the construction of
  projected lifetables.
\newblock {\em Insurance: Mathematics and economics 31}, 3 (2002), 373--393.

\bibitem{buhlmann1992stochastic}
{\sc B{\"u}hlmann, H.}
\newblock Stochastic discounting.
\newblock {\em Insurance: Mathematics and Economics 11}, 2 (1992), 113--127.

\bibitem{buhlmann2000life}
{\sc B{\"u}hlmann, H.}
\newblock Life insurance with stochastic interest rates.
\newblock In {\em Financial risk in insurance}. Springer, 2000, pp.~1--24.

\bibitem{cairns2006two}
{\sc Cairns, A.~J., Blake, D., and Dowd, K.}
\newblock A two-factor model for stochastic mortality with parameter
  uncertainty: theory and calibration.
\newblock {\em Journal of Risk and Insurance 73}, 4 (2006), 687--718.

\bibitem{clemente2021bridge}
{\sc Clemente, G.~P., Della~Corte, F., and Savelli, N.}
\newblock A bridge between local gaap and solvency ii frameworks to quantify
  capital requirement for demographic risk.
\newblock {\em Risks 9}, 10 (2021), 175.

\bibitem{clemente2022stochastic}
{\sc Clemente, G.~P., Della~Corte, F., and Savelli, N.}
\newblock A stochastic model for capital requirement assessment for mortality
  and longevity risk, focusing on idiosyncratic and trend components.
\newblock {\em Annals of Actuarial Science 16}, 3 (2022), 527--546.

\bibitem{qis2}
{\sc {Committee of European Insurance and Occupational Pensions Supervisors}}.
\newblock {\em Quantitative Impact Study 2 - Technical Specification}.
\newblock 2006.

\bibitem{delong2019fair2}
{\sc Delong, {\L}., Dhaene, J., and Barigou, K.}
\newblock Fair valuation of insurance liability cash-flow streams in continuous
  time: Applications.
\newblock {\em ASTIN Bulletin: The Journal of the IAA 49}, 2 (2019), 299--333.

\bibitem{delong2019fair1}
{\sc Delong, {\L}., Dhaene, J., and Barigou, K.}
\newblock Fair valuation of insurance liability cash-flow streams in continuous
  time: Theory.
\newblock {\em Insurance: Mathematics and Economics 88\/} (2019), 196--208.

\bibitem{dhaene2017fair}
{\sc Dhaene, J., Stassen, B., Barigou, K., Linders, D., and Chen, Z.}
\newblock Fair valuation of insurance liabilities: Merging actuarial judgement
  and market-consistency.
\newblock {\em Insurance: Mathematics and Economics 76\/} (2017), 14--27.

\bibitem{solvency2}
{\sc {European Parliament and Council}}.
\newblock {\em Directive 2009/138/EC}.
\newblock 2021.

\bibitem{GROSEN}
{\sc Grosen, A., and {Løchte Jørgensen}, P.}
\newblock Fair valuation of life insurance liabilities: The impact of interest
  rate guarantees, surrender options, and bonus policies.
\newblock {\em Insurance: Mathematics and Economics 26}, 1 (2000), 37--57.

\bibitem{hari2008longevity}
{\sc Hari, N., De~Waegenaere, A., Melenberg, B., and Nijman, T.~E.}
\newblock Longevity risk in portfolios of pension annuities.
\newblock {\em Insurance: Mathematics and Economics 42}, 2 (2008), 505--519.

\bibitem{IFRS}
{\sc {IFRS Foundation}}.
\newblock {IFRS 17 INSURANCE CONTRACTS}.
\newblock Tech. rep., 2017.

\bibitem{lee2000lee}
{\sc Lee, R.}
\newblock The lee-carter method for forecasting mortality, with various
  extensions and applications.
\newblock {\em North American actuarial journal 4}, 1 (2000), 80--91.

\bibitem{Linders}
{\sc Linders, D.}
\newblock The 3-step hedge-based valuation: fair valuation in the presence of
  systematic risks.
\newblock {\em ASTIN Bulletin 53\/} (03 2023), 1--25.

\bibitem{malamud2008market}
{\sc Malamud, S., Trubowitz, E., and W{\"u}thrich, M.~V.}
\newblock Market consistent pricing of insurance products.
\newblock {\em ASTIN Bulletin: The Journal of the IAA 38}, 2 (2008), 483--526.

\bibitem{Moehr}
{\sc Moehr, C.}
\newblock Market-consistent valuation of insurance liabilities by cost of
  capital.
\newblock {\em Astin Bulletin 41\/} (12 2010).

\bibitem{oksendal2003stochastic}
{\sc {\O}ksendal, B.}
\newblock Stochastic differential equations.
\newblock In {\em Stochastic differential equations}. Springer, 2003,
  pp.~65--84.

\bibitem{olivieri2008assessing}
{\sc Olivieri, A., and Pitacco, E.}
\newblock Assessing the cost of capital for longevity risk.
\newblock {\em Insurance: Mathematics and Economics 42}, 3 (2008), 1013--1021.

\bibitem{olivieri2015introduction}
{\sc Olivieri, A., and Pitacco, E.}
\newblock {\em Introduction to insurance mathematics: technical and financial
  features of risk transfers}.
\newblock Springer, 2015.

\bibitem{plat2011one}
{\sc Plat, R.}
\newblock One-year value-at-risk for longevity and mortality.
\newblock {\em Insurance: Mathematics and Economics 49}, 3 (2011), 462--470.

\bibitem{renshaw2006cohort}
{\sc Renshaw, A.~E., and Haberman, S.}
\newblock A cohort-based extension to the lee--carter model for mortality
  reduction factors.
\newblock {\em Insurance: Mathematics and economics 38}, 3 (2006), 556--570.

\bibitem{savelli2013risk}
{\sc Savelli, N., and Clemente, G.~P.}
\newblock A risk-theory model to assess the capital requirement for mortality
  and longevity risk.
\newblock {\em Journal of Interdisciplinary Mathematics 16}, 6 (2013),
  397--429.

\bibitem{stevens2010calculating}
{\sc Stevens, R., De~Waegenaere, A., and Melenberg, B.}
\newblock Calculating capital requirements for longevity risk in life insurance
  products: Using an internal model in line with solvency ii.
\newblock Tech. rep., Working Paper, Tilburg University, 2010.

\bibitem{wuthrich2010market}
{\sc W{\"u}thrich, M.~V., B{\"u}hlmann, H., Furrer, H., et~al.}
\newblock {\em Market-consistent actuarial valuation}, vol.~2.
\newblock Springer, 2010.

\bibitem{wuthrich2008stochastic}
{\sc W{\"u}thrich, M.~V., and Merz, M.}
\newblock {\em Stochastic claims reserving methods in insurance}.
\newblock John Wiley \& Sons, 2008.

\bibitem{wuthrich2013financial}
{\sc W{\"u}thrich, M.~V., and Merz, M.}
\newblock {\em Financial modeling, actuarial valuation and solvency in
  insurance}.
\newblock Springer, 2013.

\end{thebibliography}

\newpage

\appendix 
\section{Proofs}
\label{sec:apA}
We report here the proof of relation (\ref{eq:Idios_compact}) in Theorem \ref{Teorema_1}.
\begin{proof}
Recalling formula (\ref{vtplus1}), it is possible to rewrite it by extracting the first terms from the cross products
\begin{equation}
	\label{splitofvt}
	\begin{aligned}
		V_{t+1}=&\langle\emph{E}\left(\textbf{S}_t\big|\mathcal{T}_t\right)^\intercal, \emph{E}\left(\bm{\Lambda}_t^B\big|\mathcal{T}_t\right),U_{t+1}^{out,\;t+1}\rangle\\
		&+\langle \emph{E}\left(\textbf{S}_t\big|\mathcal{T}_t\right)^\intercal,\emph{E}\left[\left(\bm{\Lambda}_{(t+1)}^L\circ \bm{\Lambda}_{(t+1)}^B \right)\big|\mathcal{T}_t\right] , \textbf{U}_{(t+2)}^{out,\; t+1}\rangle\\
		&-\langle \emph{E}\left(\textbf{S}_t\big|\mathcal{T}_t\right)^\intercal, \emph{E}\left[\left(\bm{\Lambda}_{t+1}^L\right)\big|\mathcal{T}_t\right],U_{t+1}^{in,\;t+1}\rangle\\
		&-\langle \emph{E}\left(\textbf{S}_t\big|\mathcal{T}_t\right)^\intercal, \emph{E}\left[\left(\bm{\Lambda}_{(t+2)}^L\right)\big|\mathcal{T}_t\right],\textbf{U}_{(t+2)}^{in,\;t+1}\rangle
	\end{aligned}
\end{equation}
By definition of $\textbf{S}_{t+1}$ (see Definition \ref{def:idios}) and exploiting cross product algebra, $\hat{\mathcal{R}}_{t+1}$ can be re-written as
	\begin{equation}
		\label{splitofr}
		\begin{aligned}
			\hat{\mathcal{R}}_{t+1}=& \langle \emph{E}\left(\textbf{S}_{t+1}\big|\mathcal{T}_{t+1}\right)^\intercal, \langle \emph{E}\left(\bm{\Lambda}_{(t+1)}^L\circ\bm{\Lambda}_{(t+1)}^B\big|\mathcal{T}_{t}\right), \textbf{U}_{(t+2)}^{out,\;t+1}\rangle\\
			&-\langle \emph{E}\left(\bm{\Lambda}_{(t+1)}^L\big|\mathcal{T}_t\right),\textbf{U}_{(t+1)}^{in,\;t+1}\rangle \rangle\\
			&= \langle \emph{E}\left(\textbf{S}_{t}\circ\left(\bm{\mathds{1}}-\bm{\Lambda}_{t}^B\right)\big|\mathcal{T}_{t+1}\right)^\intercal, \langle \emph{E}\left(\bm{\Lambda}_{(t+1)}^L\circ\bm{\Lambda}_{(t+1)}^B\big|\mathcal{T}_{t}\right), \textbf{U}_{(t+2)}^{out,\;t+1}\rangle\\
			&-\langle \emph{E}\left(\bm{\Lambda}_{(t+1)}^L\big|\mathcal{T}_t\right),\textbf{U}_{(t+1)}^{in,\;t+1}\rangle \rangle
		\end{aligned}
	\end{equation}
Implementing formulas (\ref{splitofr}) and (\ref{splitofvt}) in formula (\ref{idios}), it is possible to write $CDR_{t+1}^{Idios}$ as 
	\begin{equation}
		\begin{aligned}
			CDR_{t+1}^{Idios} =& \langle\left(\emph{E}\left(\textbf{S}_t\circ\bm{\Lambda}_{t}^B\big|\mathcal{T}_t\right)-\emph{E}\left(\textbf{S}_t\circ\bm{\Lambda}_{t}^B\big|\mathcal{T}_{t+1}\right)\right)^\intercal,\left(\mathds{1}\cdot U_{t+1}^{out,\;t+1}\right)\rangle\\
			& +\langle\left(\emph{E}\left(\textbf{S}_t\circ\left(\bm{\mathds{1}}-\bm{\Lambda}_{t}^B\right)\big|\mathcal{T}_t\right)-\emph{E}\left(\textbf{S}_t\circ\left(\bm{\mathds{1}}-\bm{\Lambda}_{t}^B\right)\big|\mathcal{T}_{t+1}\right)\right)^\intercal,\left(\mathds{1}\cdot\beta_{t+1}\right)\rangle
		\end{aligned}
		\label{eq:idios_appoggio_2}
	\end{equation}
where $\beta_{t+1}$ is the generic element of the vector $\bm{\beta}_{t+1}$, defined as 
\begin{equation}
\bm{\beta}_{t+1}=\langle \emph{E}\left(\bm{\Lambda}_{(t+1)}^L\circ\bm{\Lambda}_{(t+1)}^B\big|\mathcal{T}_t\right), \textbf{U}_{(t+2)}^{out,\;t+1}\rangle-\langle \emph{E}\left(\bm{\Lambda}_{(t+1)}^L\big|\mathcal{T}_t\right),\textbf{U}_{(t+1)}^{in,\;t+1}\rangle
\end{equation} 
$\beta_{t+1}$ represents the best-estimate rate of the policyholder $k$ (we specify that all elements of the vector are equal due to assumptions of i.i.d. on survival of policyholders).\\
With simple algebra we obtain:
	\begin{equation}
		CDR_{t+1}^{Idios} = \langle\left(\emph{E}\left(\textbf{S}_t\circ\bm{\Lambda}_{t}^B\big|\mathcal{T}_t\right)-\emph{E}\left(\textbf{S}_t\circ\bm{\Lambda}_{t}^B\big|\mathcal{T}_{t+1}\right)\right)^\intercal,\left(\mathds{1}\cdot U_{t+1}^{out,\;t+1}-\mathds{1}\cdot\beta_{t+1}\right)\rangle
	\end{equation}
Considering linearity of the expected value formula (\ref{eq:Idios_compact}) is proved.
\end{proof}

\begin{proof} Here we prove formulas (\ref{eq:sigma_y}).\\
We introduce the hedgeable filtration (see \cite{malamud2008market}) $\mathbb{H}=\left(\mathcal{H}_t\right)_t=0,...,n-1$ and
\begin{equation}
	\mathcal{H}_{t+1}=\sigma\left\{\mathcal{T}_t,\mathcal{G}_{t+1}\right\}
\end{equation}
We observe
\begin{equation}
	\begin{aligned}
&Var\left(CDR_{t+1}^{Idios}\big|\mathcal{F}_t\right)\\
&=\emph{E}\left[Var\left(CDR_{t+1}^{Idios}\big|\mathcal{H}_{t+1}\right)\big|\mathcal{F}_t\right]+Var\left[\emph{E}\left(CDR_{t+1}^{Idios}\big|\mathcal{H}_{t+1}\right)\big|\mathcal{F}_t\right]\\
&=\emph{E}\left[Var\left(CDR_{t+1}^{Idios}\big|\mathcal{H}_{t+1}\right)\big|\mathcal{F}_t\right]\geq0
	\end{aligned}
\end{equation}
	Switching from a matrix representation to an extended one, formula (\ref{eq:Idios_compact}) can be represented as 
	\begin{equation}
CDR_{t+1}^{Idios} = \sum_{k=0}^{l_0}s_{k,t}\cdot\left[\emph{E}\left(\Lambda_{k,t}^B\big|\mathcal{T}_t\right)-\emph{E}\left(\Lambda_{k,t}^B\big|\mathcal{T}_{t+1}\right)\right]\cdot \eta_{t+1} 
	\end{equation}
	Let define as $Z_t=\langle \left(\textbf{S}_t\right)^\intercal,  \bm{\Lambda}_{t}^B\rangle$ the total sums insured of occurred deaths at time $t$ and $Z_{k,t}$ the sum insured of occurred death of the single $k$-th policyholder.
	Since  the generic element of $\bm{\Lambda}_{t}^B$ is distributed as a Bernoulli r.v. with parameter $q_{x+t}$, the moment generating function (mgf) $M_{\lambda_{k,t}^B}(m)$ is
	\begin{equation}
		M_{\lambda_{k,t}^B}(m)=q_{x+t}\cdot e^{m}+(1-q_{x+t})
	\end{equation}
	Considering the policyholders' sums insured, we have:	\begin{equation}
		M_{Z_{k,t}}(m)=q_{x+t}\cdot e^{m \cdot s_{k,t}}+(1-q_{x+t})
	\end{equation}
	Considering that elements of $\bm{\Lambda}_{t}^B$ are identical distributed, we have: 
	\begin{equation}
		M_{Z_{t}}(m)=\left(q_{x+t}\cdot e^{m \cdot s_{k,t}}+(1-q_{x+t})\right)^{l_t}
	\end{equation}
	Therefore, the cumulant generating function is defined as:
	\begin{equation}
		\Psi_{Z_{t}}(m)={l_t}\cdot ln\left(q_{x+t}\cdot e^{m \cdot s_{k,t}}+(1-q_{x+t})\right)
	\end{equation}
	Hence, the variance is defined as:
	\begin{equation}
		Var(Z_t)=l_{t}\cdot\ q_{x+t}\cdot(1-q_{x+t})\cdot \bar{\textbf{S}}_t^2
	\end{equation}
	where $\bar{\textbf{S}}_t^j$ denotes the $j$-raw moment of the $\mathcal{F}_t$-adaptable vector $\textbf{S}_t$. 
\end{proof}

\end{document}